\documentclass{article}
\usepackage{graphicx} 
\usepackage{biblatex}
\usepackage{xcolor}
\usepackage{amsmath}
\usepackage{authblk}

\title{SITH: A Quantum-Chemical Framework for Predicting Bond Destabilization in Stretched Molecules}

\author[1,2,4]{Daniel Sucerquia}

\author[5]{Mikaela Farrugia}

\author[3]{Andreas Dreuw}

\author[1,2,3]{Frauke Gr\"ater\thanks{Corresponding author: graeter@mpip-mainz.mpg.de}}

\affil[1]{Heidelberg Institute for Theoretical Studies, Heidelberg 69118, Germany}
\affil[2]{Max Planck Institute for Polymer Research, Mainz 55128, Germany}
\affil[3]{Interdisciplinary Center for Scientific Computing, Heidelberg University, Heidelberg 69120, Germany}
\affil[4]{Department of Physics and Astronomy, University of Heidelberg, Heidelberg 69120, Germany}
\affil[5]{Department of Chemistry \& Biochemistry, University of Notre Dame, Notre Dame, Indiana 46556, United States}
\begin{document}

\maketitle

\abstract{
Mechanical forces  can selectively destabilize chemical bonds of molecular systems, particularly in biological and synthetic polymers. While experimental and theoretical methods have advanced our understanding of mechanochemical processes, predicting where energy concentrates within a molecule remains a significant challenge. To address this, we introduce SITH (Splitting Intramolecular Tension due to stretcHing), a novel method that decomposes the total electronic energy change of a stretched molecule into contributions from individual degrees of freedom ---such as bond lengths, angles, and dihedrals--- using numerical integration of the work-energy theorem. Unlike previous approaches that rely on harmonic approximations, SITH provides high accuracy and robustness to study the distribution of energies of stretched molecules up to a first bond cleavage. Although SITH uses 3N-6 degrees of freedom for the energy decomposition, we show that it can work even for ring structures like prolines. We apply SITH to a dataset of tripeptides and demonstrate that glycine and proline exhibit significantly different energy distributions in their C$_\alpha$–C backbone bonds under tension: glycine stores more energy, making it more prone to rupture, while proline has the opposite behaviour. These findings reveal intrinsic differences in mechanochemical susceptibility across amino acids, offering more accurate predictions of bond rupture in proteins, and similarly in other (bio)polymers. SITH thus provides a powerful, interpretable tool for understanding energy distribution at the quantum level, with possible implications in mechanochemistry and force field validation.
}

\section{Introduction}

Mechanical force, being it an externally applied stretching force or an internal strain, stretches molecules and eventually breaks them apart. An external force can increase the probability of rupture of specific bonds~\cite{stauch_advances_2016, liu_review_2022}, or even can strengthen what are called catch bonds~\cite{thomas_biophysics_2008, mulla_weak_2022}. Mechanochemistry has been extensively explored in synthetic systems such as polymers~\cite{hsu_primitive_2018, hsu_clustering_2019} and strained ring structures~\cite{rowe_chemical_2023, chow_proline_2018}. Molecules in living organisms are also constantly being pushed and pulled, and it has recently been recognized that chemical bonds in protein materials also break at physiological levels of stretching forces ~\cite{zapp_mechanoradicals_2020, rennekamp_collagen_2023}.

These effects have been studied at the molecular level in the laboratory with techniques such as single-molecule force spectroscopy (SMFS) ~\cite{neuman_single-molecule_2008}, sonication combined, for example, with fluorescent mechano-sensors~\cite{cravotto_mechanochemical_2013} and material stretching coupled to mechanoradical detection ~\cite{rennekamp_collagen_2023}. Several theoretical methods, mostly based on quantum chemical calculations, have given detailed insights into the underlying processes, the scissile bonds, preferred reaction pathways, and force-rate relationships~\cite{ribas-arino_covalent_2012, stauch_advances_2016, hickenboth_biasing_2007}. An interesting question is how the mechanical force distributes through a complex molecular system, thereby causing a specific mechanochemical reaction, such as bond rupture, to occur. How does an external force differently destabilize bonds in a molecule? Which is the weakest bond in a stretched molecule, namely, the most probable to break first? Answering these questions is non-trivial as quantum mechanical descriptions cannot, straightforwardly, be decomposed into force or energy contributions in individual bonds or other degrees of freedom.

In a classical bead-spring-like model instead, used for example in molecular mechanics force fields, the total energy is decomposed into individual bonded and non-bonded terms, such as terms due to bond or angle stretching. Using such a model, Hsu and Kremer studied how polymer melts respond to strong deformations and where forces concentrate in such highly entangled system~\cite{hsu_primitive_2018, hsu_clustering_2019}. In such homopolymers, bond types are the same across all monomers, and what matters is their interactions and entanglement. For heteropolymers such as proteins, crosslinked polymers and systems with mechanophores, distinct bond types can be destabilized by stretching very differently. Considering each bond as a  Morse potential, a force deforms the bond and lowers the energy barrier towards rupture by adding work  to the potential energy along the bond direction, as described by the Bell-Evans model~\cite{bell_models_1978}. Rennekamp {\it et al.} used this approximation in a simulation framework, called KIMMDY, which couples molecular dynamics and kinetic Monte Carlo,  to simulate bond ruptures of pulled molecules~\cite{rennekamp_collagen_2023, rennekamp_hybrid_2020, hartmann_kimmdy_2025}.
Thus, decomposing the energy of a stretched molecule into energy components stored in the individual bonds, or more generally in the different molecular degrees of freedom,  is not only of conceptual interest but also has practical applications. It can serve as a basis for predicting rates and thereby the dominant mechanochemical reaction pathways.

\begin{figure}
    \centering
    \includegraphics[width=\textwidth]{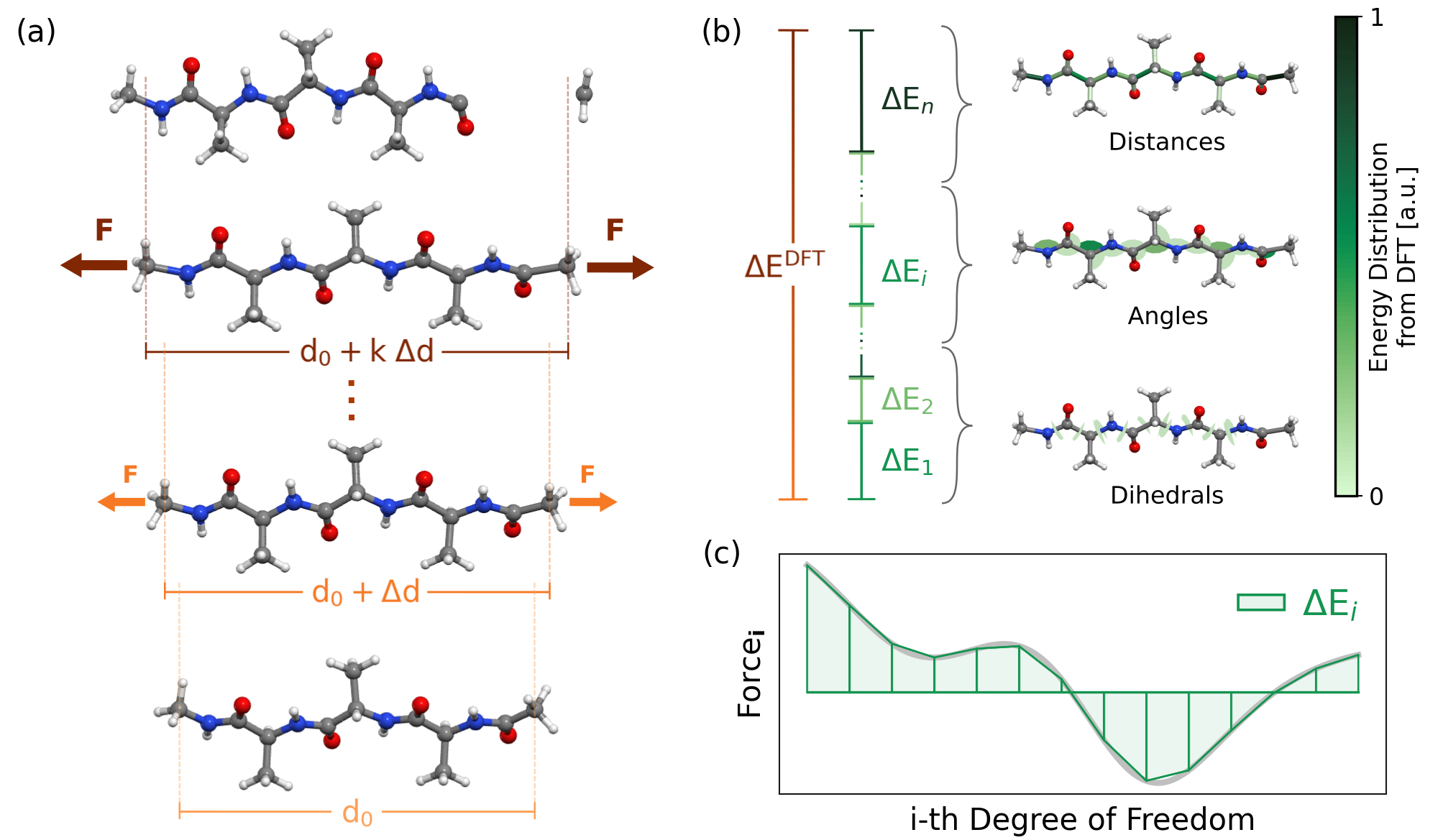}
    \caption{Illustration of SITH stretching and energy distribution. a) stretching process, starting from the optimized configuration (bottom), we increase the end to end distance by a value $\Delta$d and reoptimize with this distance constrained, which is equivalent to a stretching force pulling from the ends. The process is repeated k times, until k+1 produces a rupture (top). b) SITH describes a decomposition of the total energy obtained with DFT in terms of energies of n degrees of freedom ($\Delta E_i$) based on the DFT calculation during the stretching process. The degrees of freedom are distances, angles, and dihedrals that are colored according to the magnitude of the corresponding change of the energy. c) The value of the energy $\Delta E_i$ comes from a numerical integration of the force in the $i$-th degree of freedom, as it changes during the stretching process.}
    \label{fig:SithIllustration}
\end{figure}

Density functional theory (DFT) is widely used in mechanochemistry to describe the changes in stability of molecules under external forces. For example, the DFT-based COGEF method assesses the energies in a molecule all the way to rupture ~\cite{beyer_mechanical_2000}. This method mimics the effect of an external force pulling two atoms in opposite directions by increasing the distance between these atoms and optimizing the structure while keeping the new distance constrained. The stretching is repeated until bond rupture is observed (Fig.~\ref{fig:SithIllustration}a). Here and throughout, a cleaved molecule is defined as the structure for which all the bond lengths relax except for the broken bond. In other words, this method estimates the energy required to break a molecule mechanically at zero temperature into two parts. This approach yields the energy for rupture along with the molecular deformations before rupture, but cannot identify the bonds where the externally applied force concentrates. That is, it does not give a description of which part of a molecule loads more or less energy, namely, which bond is the most susceptible to break.

To address the  challenge of predicting which parts of a molecule are less and which parts are more destabilized by an external stretching force, Stauch and Dreuw proposed a method, JEDI, that describes the energy stored in redundant internal coordinates (RIC) (all bond distances, angles, and dihedrals) of an elongated molecule ~\cite{stauch_quantitative_2014, stauch_quantum_2017}. This method consists of a second order expansion of the energy, which uses the RIC Hessian to predict the energy cumulated in each one of the degrees of freedom,  expecting that the sum of the individual energies approximates the total DFT energy (Fig.~\ref{fig:SithIllustration}b). This can be summarized as

\begin{eqnarray}\label{eq:jedi}
    \Delta_0^{k}E_{DFT} \approx \Delta_0^{k} E_{harm} & = & \sum_i \Delta_0^{k}E_i \\
                             & = & \sum_i \sum_j\frac{\partial^2 V}{\partial q_i \partial q_j}\bigg|_{{\{q\}}^0} \hspace{2mm}\Delta_0^k q_i \hspace{2mm}\Delta_0^k q_j ~,
\end{eqnarray}

\noindent
where $V$ is the potential energy, $q_i$ is the value of the $i$-th RIC, and $\Delta_0^{k}$ is the difference operation of the $k$-th stretched configuration in respect to the optimized one, which we will use $\Delta$ for notational simplicity, unless other indices are stated. In this way, it is possible to assess the energy associated with each degree of freedom using only three components: the Hessian matrix of the optimized configuration, the optimized geometry, and the stretched structure of interest. An advantage of JEDI is that, because of the use of RICs, it can be used to describe molecules with highly connected atoms, like benzenes or rings in general. However, it is limited to stretched configurations close to the optimized geometry and therefore cannot be used to study energy distributions near bond cleavage.

Avdoshenko {\it et al.} showed that, in mechanical equilibrium with an external force, the total work of a force deforming a molecule from a state $0$ to a state $k$ can be described in terms of a generalized force defined as the partial derivative of the potential energy in respect to a set of variables~\cite{avdoshenko_calculation_2014}. Those variables have to form a complete and independent set that describes the geometry of the system, removing symmetries of translations and rotations. This means that the distribution of forces is obtained for a limited set of 3N-6 degrees of freedom. The main application of this method has been to analyze distribution of forces. To the best of our knowledge, it has not been used to study bond reactivity.

Inspired by the methods described before, here we propose SITH,  Splitting Intramolecular Tension due to strecHing, to obtain a more accurate and interpretable method for assessing the distribution of molecular energies at DFT level of accuracy. SITH calculates the distribution of energies in different degrees of freedom from the work, by numerically integrating the force without requiring a harmonic approximation, unlike JEDI. In section~\ref{sec:methods}, we explain in detail the basis of our method. Then, we proceed first to show an application to trialanine. Subsequently, we discuss the challenge of describing the distribution of energies in the backbone of prolines, amino acids featuring 5-rings, and show that our SITH analysis does not depend on the set of degrees of freedom that we choose, given that the backbone forms the force-bearing scaffold in this case. Then, we continue with a complete study of tripeptides with different amino acids to investigate which are primarily weakened under an external force. We find that glycine and proline present the least and most stable C$_\alpha$-C bonds upon stretching, respectively. The rest of the amino acids do not present statistically relevant differences to be highlighted, implying that the nature of the sidechains do not alter the distribution of energies along the backbone. Finally, we discuss the applicability and accuracy of SITH.

\section{Methods}\label{sec:methods}

The SITH analysis is inspired by previous methods to obtain a decomposition of the total change of the DFT energy into changes of energies associated with certain degrees of freedom. The set of degrees of freedom chosen is arbitrary, but it has to be a complete linearly independent set of $n=3N-6$ variables that can describe all states of the molecule, where $N$ is the number of atoms. We choose a Z-matrix as a rule of thumb, being aware that there are different ways to construct z-matrices for the same system~\cite{gordon_approximate_1968}. Our advice is to choose one that contains most of the degrees of freedom of interest (distances, angles, and dihedrals defined by bonded atoms), given that in this way we can guarantee some interpretability.

The first step of SITH is to pull the molecule of interest using the same procedure as COGEF, generating then a set of $k$ configurations optimized while constraining the distance between two atoms (typically in the terminal groups), which is increased with every step (Fig. ~\ref{fig:SithIllustration}a). With the set of generated configurations, we reconstruct the energy distribution also aimed by JEDI (Fig. ~\ref{fig:SithIllustration}b), but instead of using a harmonic approximation, we obtain the energies stored in each degree of freedom from the integral of the work-energy theorem, which we solve numerically using the trapezoid rule~\cite{epperson_introduction_2013} (Fig.~\ref{fig:SithIllustration}c). In contrast to JEDI, SITH does not require a harmonic approximation for these degrees of freedom. In other words, the energy associated with the i-th degree of freedom can be defined as

\begin{eqnarray}
    \Delta E_i & = & - \int_{0}^{k} F_i dq_i \\
               & \approx & - \sum_{j=0}^{k} \frac{F^j_i + F^{j+1}_i}{2} \Delta_j^{j+1} q_i~, \label{eq:numerical_integration}
\end{eqnarray}

\noindent
where $\Delta_j^{j+1} q_i$ is the change of the value of the $i$-th degree of freedom between two consecutive configurations in the COGEF path. $F^j_i$ is the force associated to the i-th degree of freedom for the j-th configuration in the path, which can be computed from the DFT energy via the generalized force expression proposed by Avdoshenko {\it et al.}~\cite{avdoshenko_calculation_2014},

\begin{equation}\label{eq:force_i}
    F^j_i = -\frac{\partial E^{DFT}}{\partial q_i}\bigg|_{\{q\}^j} ~.
\end{equation}

We point out that in the original paper of Avdoshenko {\it et al.}, the definition of the generalized force does not include the negative sign, which is just a different convention considering that $F^j_i$ can be seen as the external force changing the structure in the molecule or the internal force opposing this change; one is the negative of the other. The accuracy of both JEDI and SITH can be tested by summing up all the contributions of the different degrees of freedom, and comparing this sum to the total change of energy obtained from DFT (Fig.~\ref{fig:SithIllustration}b),

\begin{equation}\label{eq:total_energy}
    \Delta E^{total} = \Delta E^{DFT} \approx \sum_i \Delta E_i~.  
\end{equation}

The error of the trapezoidal rule used in equation~\ref{eq:numerical_integration} is defined as~\cite{epperson_introduction_2013}

\begin{equation}\label{eq:error}
    Err(\Delta E_i) = \sum_{j=0}^{k} \frac{(\Delta_j^{j+1} q_i)^2}{12} \left[\frac{\partial^2 E^{DFT}}{\partial q_i^2}\bigg|_{\{q\}^{j+1}} - \frac{ \partial ^2 E^{DFT}}{\partial q_{i}^2}\bigg|_{\{q\}^j}\right]~.
\end{equation}

To ensure high accuracy in the numerical integration, the variations in the degrees of freedom ---$\Delta_j^{j+1} q_i$ in equation \ref{eq:numerical_integration}--- are kept as small as possible. Naturally, this introduces a trade-off between the number of intermediate configurations and the computational cost required to obtain optimized configurations and energy derivatives using DFT.

Due to mechanical stretching, the energy stored in the bonds increases, leading to a reduction in the activation barrier for bond rupture. Within the framework of the Bell–Evans model~\cite{bell_models_1978}, we evaluate the modified activation energy barrier as (see discussion about energy barriers in SI)

\begin{equation}\label{eq:newbarrier}
\begin{split}
    D' = & \hspace{2mm} D\sqrt{{\frac{-4\sqrt{ D\Delta E_i} +  4\Delta E_i + D}{D}}} \hspace{3mm}+ \\
         & \hspace{2mm} 4 (\Delta E_i - \sqrt{D \Delta E_i}) \hspace{2mm}\textbf{tanh}^{-1}\left(\sqrt{\frac{-4\sqrt{D\Delta E_i} + 4\Delta E_i +D}{D}}\right)~,
\end{split}
\end{equation}

\noindent
where $D$ is the bond dissociation energy of the $i$-th bond.

In principle, the energy of each degree of freedom, $E_i$, depends on the values of the other degrees of freedom (equation \ref{eq:force_i}). However, we can assume as an approximation that the DFT energy, $E^{DFT}(\{q\})$, can be expanded as the sum of independent energies $E_i(q_i)$. This is,

\begin{equation}\label{eq:interactions}
    E^{DFT}(\{q\}) \approx \sum_i E_i(q_i)~.
\end{equation}

From this point of view, we interpret the SITH analysis as a model of a molecular system in terms of fundamental interactions. This idea is also used in molecular dynamics. For example, in classical force fields, the interaction $E_i(q_i)$ of the bond distances is approximated to a harmonic potential that depends only on the interatomic distance of the atoms forming the bond ($q_i$). In this way, the prediction of SITH is analogous to the bond interactions of classical force fields, as we discuss in  section~\ref{sec:trialanine}.

In practice, the user of SITH can obtain the forces of equation~\ref{eq:force_i} by applying the transformation of the Cartesian forces into the base of internal $3N-6$ coordinates, using the same formalism of Stauch \& Dreuw~\cite{stauch_quantitative_2014} or using the method of Avdoschenko {\it et al.}~\cite{avdoshenko_calculation_2014}. For the practical application of SITH, we created a package that includes the workflow to do all the necessary steps to complete the stretching, energy decomposition, and analysis shown in the next section (see code availability). The SITH software uses gaussian09 for all quantum chemical calculations~\cite{frisch_gaussian_2009}. Due to its modular structure, SITH can be expanded to work with other quantum chemistry packages. We use BMK/6-31+g(2df,p) as quantum level of accuracy~\cite{boese_development_2004}.

\section{Results}

\subsection{Establishing SITH for Trialanine as Model System}\label{sec:trialanine}

\begin{figure}
    \centering
    \includegraphics[width=\textwidth]{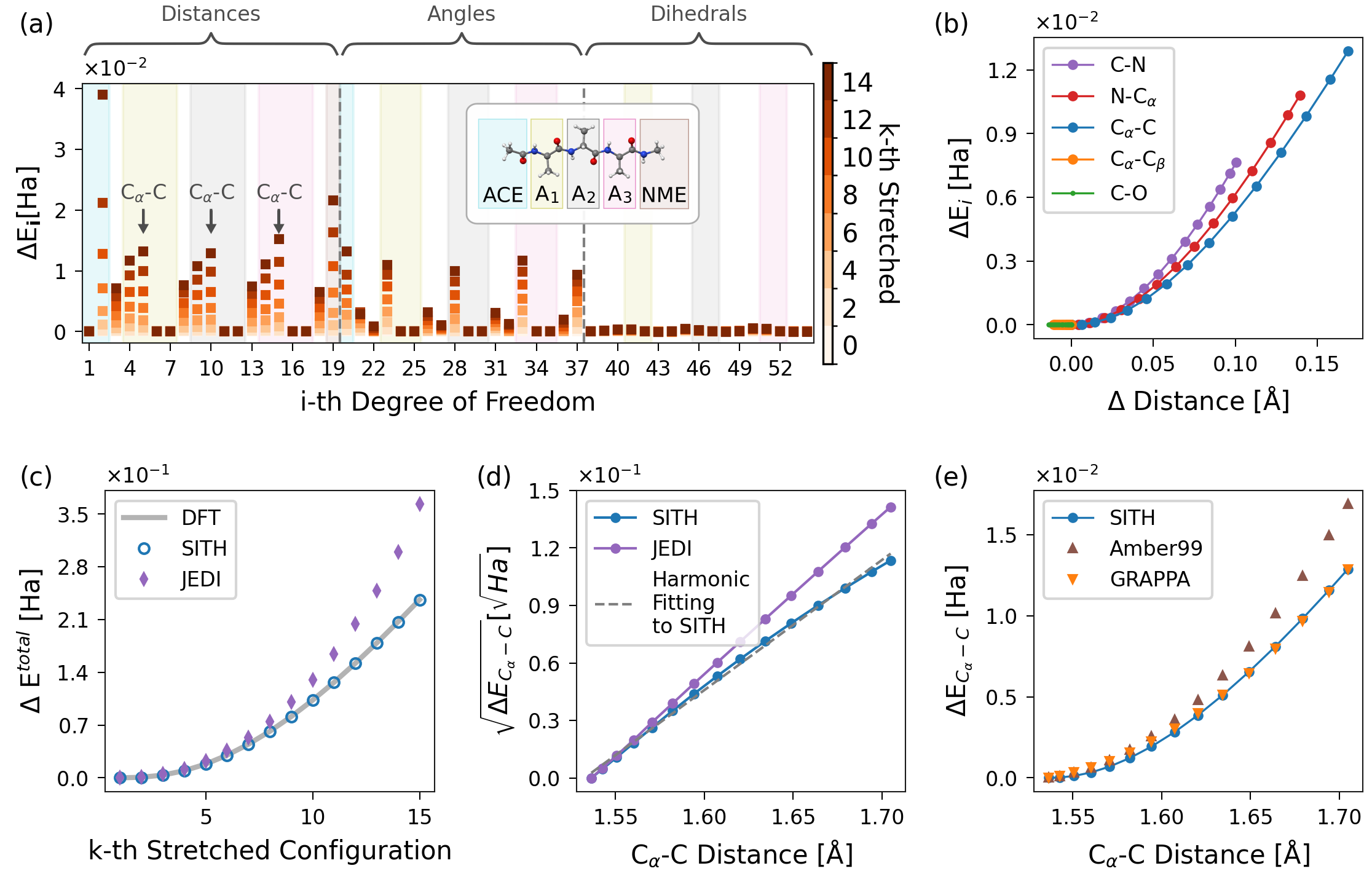}
    \caption{SITH energy decomposition for tri-alanine. a) Energy stored in each degree of freedom as the stretching increases. The degrees of freedom are grouped in distances, angles, and dihedrals, with background colors indicating the corresponding residue: ACE and NME capping groups and the three alanines (white: atoms of the two adjacent residues are involved). b) Energy stored in the bonds of the central alanine of the tri-peptide (A$_2$). C-N corresponds to the bond between C atom of alanine 1~(A$_1$) and N atom in alanine 2~(A$_2$). c) Total change of energy predicted by SITH and JEDI compared with the expected energy from DFT. d) Root squared energy stored in the C$_\alpha$-C distance according to the prediction by SITH, JEDI, and a harmonic fit to the SITH data. e) Energy stored in the C$_\alpha$-C distance according to SITH (based on quantum chemical calculations) compared to the bond energy predicted by classical force fields Amber99 and GRAPPA.}
    \label{fig:trialanine}
\end{figure}

While SITH is a method generally applicable to any stretched (macro)molecule for which the force-bearing scaffold contains less than 3N-6 degrees of freedom, here we focus on its application to proteins. We start from small linear systems which are devoid of effects from topology and entanglement and that can be reliably treated with DFT.  As a first simple model, we consider a peptide formed by three alanines, such that also the chemical buildup of each monomer --- and thus its effect on the energy distribution along the backbone --- is the same. We start our simulation from the optimized configuration and then increase the distance between the CH3 atoms of the NME and ACE capping groups until producing bond cleavage (Fig.\ref{fig:SithIllustration}a and Fig.~\ref{fig:trialanine}a). Figure~\ref{fig:trialanine}a shows the distribution of energies in the different degrees of freedom of the trialanine as the stretching increases, with each residue and capping group highlighted by another background shade. The error stemming from numerical integration (equation~\ref{eq:error}) is on the order of $10^{-9}$ Ha and thus completely negligible. As expected, energies per degree of freedom increase during stretching. We notice that the distances, i.e. chemical bonds, store more energy than the angles, while the dihedrals mostly do not store any energy. The latter is probably a consequence of starting the stretching from an optimized extended configuration. 

We also notice that the degree of freedom storing the largest amount of energy during the stretching process corresponds to the first bond to break, which is the CH3-C bond in the ACE capping group. This is evidence that the distribution of energies computed by SITH is directly related to the rupture propensity (see also the discussion about the barrier change in SI and Fig. S8). However, the capping groups are only included here in order to work with a small system, and in the analysis that follows, we focus on the amino acids in between.

We observe some variations, even though rather minor, in the stored energies across the three central alanines.  This implies that the distribution of energies in a given amino acid depends to some extent on the neighboring residues. This probably happens because the electronic distribution differs depending on the environment (we expand more on this in Section \ref{sec:dataset}). To have a notion of the degree of influence of the capping groups, we applied SITH to a peptide formed by five alanines (Fig. S1). We see that the three alanines in the center yield the same energies per degree of freedom, while the two closest to the capping groups are slightly different. We conclude that a residue influences up to its first neighbor's distribution of energies. This implies that, to analyze the energies of a given amino acid, it is important to also consider the first neighbors. Moreover, it is important to avoid the capping groups being its first neighbors to avoid artifacts. In other words, for modeling the energy distribution along a linear peptide chain, it is sufficient to consider peptides with three amino acids and to focus on the amino acid in the middle.

Among the distances in the alanine, the C$_\alpha$-C distance dominates the storage of the energy, followed by the C$_\alpha$-N distance and the C-N distance (Fig.~\ref{fig:trialanine}a, b). Thus, the peptide bond with its partial double bond character is less extended and less destabilized than the single bonds in a peptide backbone.
As expected, some degrees of freedom are not affected by the external force. The analysis of the distribution of energies in the distances of the alanine in the middle shows that the side chains do not store energy (for example, C$_\alpha$-C$_{\beta}$ in Fig.~\ref{fig:trialanine}b). However, the value of their distances may vary because the external force changes the potential energy surface, such that the value of the minimum of energy of this degree of freedom changes its location, but not its energy.

In contrast to JEDI, SITH shows to be highly accurate compared to the expected change of total energy (see equation~\ref{eq:total_energy} and Fig.~\ref{fig:trialanine}c). While the sum of the energy components of the SITH decomposition fits almost perfectly to the expected DFT total change of energy, JEDI overestimates this value. This is because, as the main assumption of JEDI is a harmonic approximation, it only works for low stretching that keeps the altered configuration close to the minimum. Indeed, the blue and the dashed gray curves in Fig.~\ref{fig:trialanine}d show that the energy associated with the C$_\alpha$-C distance slightly deviates from a completely harmonic behavior. However, the main source of error of JEDI is that it predicts stiffer interactions than they really are (slope of the curves in Fig. ~\ref{fig:trialanine}d). For the specific case shown in  Figure~\ref{fig:trialanine}d, SITH predicts a spring constant of 0.23 Ha/\AA$^2$, while JEDI predicts a spring constant of 0.35 Ha/\AA$^2$.

As we describe in the methods (section~\ref{sec:methods}), the approach to decompose energies in terms of independent functions is analogous to classical molecular mechanics force fields. We show in Fig.~\ref{fig:trialanine}e how our SITH prediction of the energy on a specific bond compares with the  values used for two force fields: Amber99~\cite{cornell_second_1995} and GRAPPA, a machine-learned force field trained on DFT energies and forces~\cite{seute_grappa_2025}. We observe that both force fields follow the SITH result quite closely, in particular up to intermediate bond elongations, for which they have been parametrized. Interestingly, overall GRAPPA fits better to SITH than Amber99, for this bond and also for other degrees of freedom (Fig. S2). This makes sense because GRAPPA parametrization comes exclusively from electronic structure calculations, while Amber99 also takes experimental observables at room temperature into account. We show with this comparison that SITH predicts reasonable values of the bond interactions and could even be used as a benchmark of classical force fields, specially for studying the effects of forces that alter the molecular structure far from the equilibrium.    

\subsection{SITH can Handle Ring Structures: Proline}

One of the conditions to work with SITH is to select 3N-6 internal degrees of freedom. For this purpose, we suggest using Z-matrices and,  as necessary condition, including the relevant degrees of freedom  of the molecule (defined by bonded atoms), which is also called the force-bearing scaffold. But using a Z-matrix implies adding only one bond per new atom in the construction of the molecule. By definition, then, molecules with rings (like prolines), cannot be defined using all the bonds of the ring, and at least one bond has to be left out. We wonder, then, how much the distribution of energies varies when we choose a different set of degrees of freedom. Avdoshenko {\it et al.} addressed this question for other systems and showed that the distribution of forces (for the purpose of this paper, also the distribution of energies) significantly depends on the set of degrees of freedom chosen for the analysis~\cite{avdoshenko_calculation_2014}. Here, we investigate how sensitive the results of SITH are with respect to the choice of bonds within the proline ring, given that the C$_\alpha$-N bond is always included as part of the force-bearing scaffold.

\begin{figure}
    \centering
    \includegraphics[width=\textwidth]{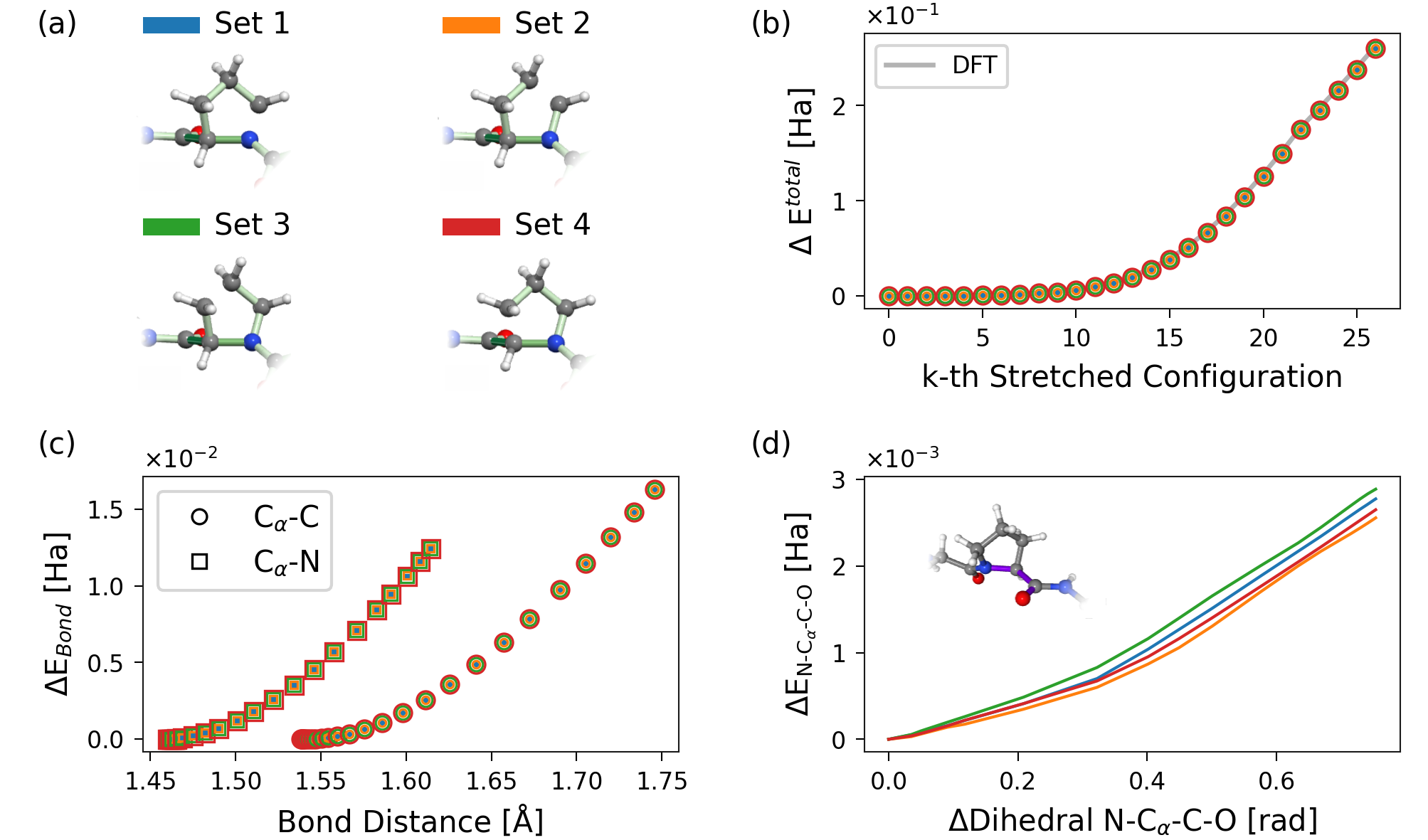}
    \caption{SITH energy analysis of proline. a) Set of degrees of freedom used in the analysis. Each set leaves out one bond of the ring to avoid more than $3N - 6$ degrees of freedom, where $N$ is the number of atoms. b) Total change of energy in the stretching predicted by SITH using the different sets and compared with the expected DFT change of energy. c) Energy in the C$_\alpha$-C and the C$_\alpha$-N distances computed by SITH using the 4 sets. d) Energy stored in the dihedral angle N-C$_\alpha$-C-O, which is the degree of freedom that differs the most when using different sets.}
    \label{fig:proline}
\end{figure}

To address this issue, we study a peptide formed by three amino acids: glycine, proline, and methionine (GPM). We locate the proline in the middle, such that the capping groups do not bias our analysis. Figure~\ref{fig:proline}a shows the different set of bonds, for which a different bond of the ring is neglected. Although we change the set of degrees of freedom, the stretching path that we use in the four cases is the same. When we compare the total energy predicted by SITH using the four different sets (equation~\ref{eq:total_energy}), we see a perfect agreement among the four sets and with respect to the expected total energy computed from DFT (Fig.~\ref{fig:proline}b). This means that the energy redistributes in the degrees of freedom of the different sets in such a way that the total energy is preserved.

Then, we search the degrees of freedom common to the four sets that differ in the energy distribution. We first focus on the energies of the C$_\alpha$-C and the C$_\alpha$-N distances, because they are the most relevant bonds of the backbone, as shown in the case of the tri-alanine (Fig.~\ref{fig:trialanine}a). We observe that they do not differ when changing the set of degrees of freedom (Fig.~\ref{fig:proline}c); all the sets agree on the value of the energies stored in these bonds of the backbone. The degree of freedom showing the largest variation is the N-C$_\alpha$-C-O dihedral angle (Fig.~\ref{fig:proline}d). However, we notice that the energy stored in this dihedral is one order of magnitude lower than the energy stored in the bonds (Figure~\ref{fig:proline}c). Moreover, even for this case, the tendency for the four cases is the same, and the maximum variation of the predicted energy among the four sets is 3.3x10$^{-4}$ Ha; one order of magnitude lower than the value of the energy itself.

This indicates that we can use SITH to study the distribution of energies in the bonds of the backbone of peptides containing proline independently of which bond of the ring we leave out of the analysis. This can be seen as a contradiction with the results shown by Avdoshenko {\it et al.} A possible reason for that is that they compare completely random sets of degrees of freedom, which could then lose interpretability and comparability. In contrast, we keep those degrees of freedom in our sets that are relevant for the analysis and that have a physical interpretation. Of course, as soon as the C$_\alpha$-N bond of proline ---as part of the force-bearing scaffold--- is left out, an unphysical energy distribution could be obtained. Therefore, we only use interactions through bonds in the definition of our sets, which perfectly preserve, as we show here, the values of the predicted energies.

\subsection{Energy Distribution in Tripeptides}\label{sec:dataset}

After having established SITH as a method to faithfully decompose energies into the relevant degrees of freedom, we now move one step up in complexity and compare the energy distribution across different types of amino acids. Here, we still leave the influence of topology or entanglement aside and consider linear chains of amino acids only. In this way, we now can ask to what extent amino acids differ in how much energy they distribute into their backbone degrees of freedom, thereby stabilizing or destabilizing the backbone locally, solely based on the different local chemical environment. 

\begin{figure}
    \centering
    \includegraphics[width=\textwidth]{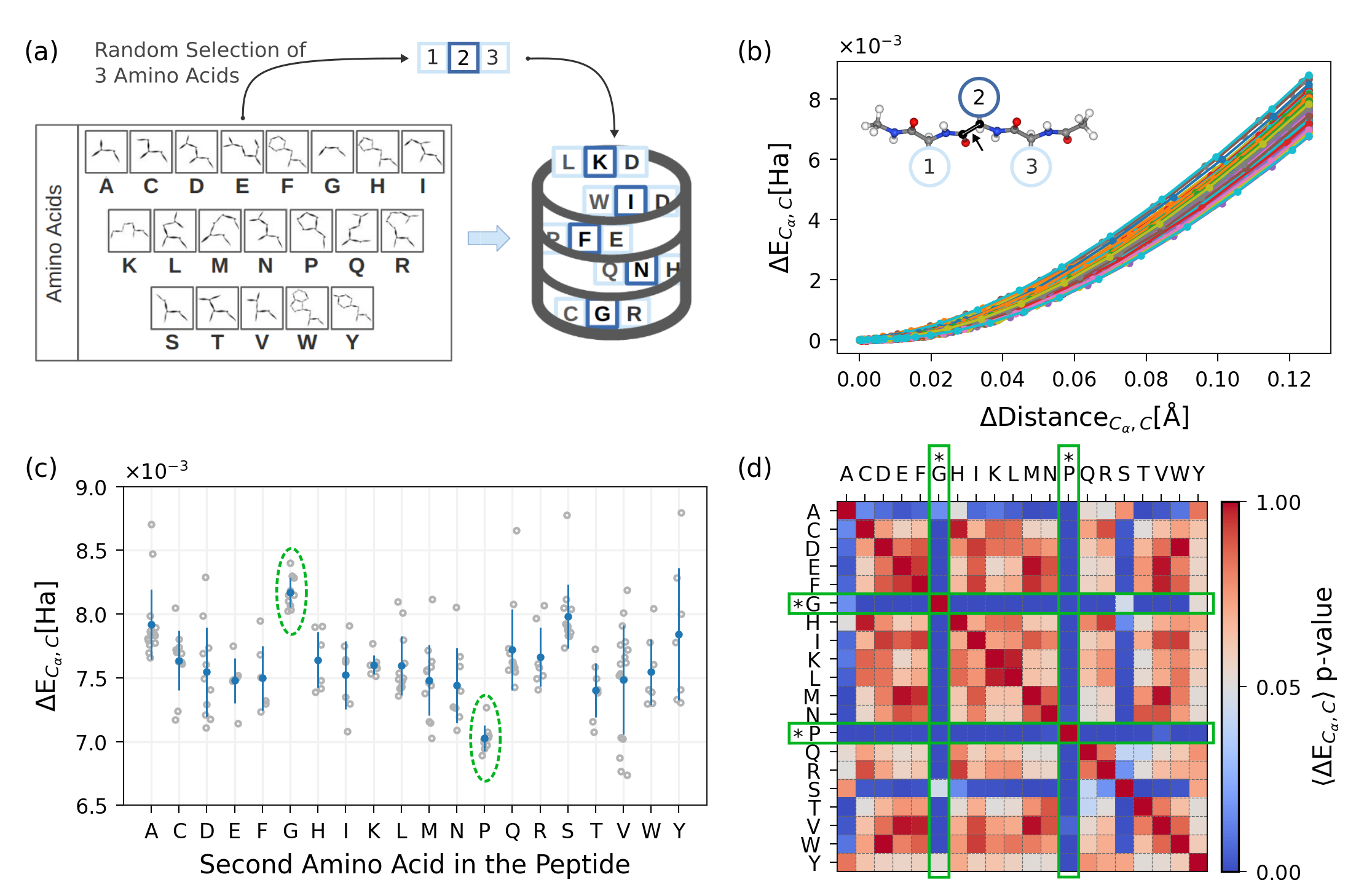}
    \caption{SITH analysis for a dataset of tri-peptides. a) We create a dataset of tripeptides by selecting random sets of three amino acids from the 20 natural amino acids. Here, we show the schematic structure of the amino acids and the associated one-letter code (see Table S1 for a complete description). b) Energy stored in the C$_\alpha$-C of the central amino acid in the peptides as a function of the change of the distance with respect to the equilibrium distance. Each curve corresponds to a different peptide. c) Distribution of energies stored in the C$_\alpha$-C distance for the peptides in the dataset after stretching the bond 0.125{\AA} from the equilibrium distance, shown as a function of the amino acid in the middle. The gray points are energies of the different peptides with the different combinations of amino acids in the first and third positions. The blue dots are the mean values, and the bars are the standard deviations. d) p-values of the energies in the C$_\alpha$-C between every pair of amino acids in the second position of the peptide. The colormap is a divergent scheme where white is the threshold of an accepted p-value of 0.05. Along with the f-statistics test (Fig. S4), we find which distributions statistically differ from the others. In c) and d), glycine and proline are highlighted because their distributions are significantly shifted from the distribution of the other amino acids.}
    \label{fig:dataset}
\end{figure}

As shown before (Section~\ref{sec:trialanine}), considering a tripeptide and placing the amino acid of interest in the middle is sufficient. If we were to consider all possible combinations of three amino acids, we would have to implement our SITH analysis in $20^3$ peptides, which is computationally expensive and unnecessary to undertake. As an alternative, we create a dataset of 190 tripeptides formed by random selection of amino acids (Fig.~\ref{fig:dataset}a). Figure S3 shows the amount of amino acids at each position of the peptides in the dataset (see also data availability for more details). We stretch each peptide and apply the SITH analysis as described above. 

Figure \ref{fig:dataset}b shows the energy of the C$_\alpha$-C bond of the amino acid in the center of each peptide when increasing the interatomic distance {0.125\AA} from the equilibrium. The first point to notice is that the different peptides rapidly show a variation of 2x10$^{-2}$ Ha in the amount of energy stored at this stretching distance. Although the deviation is small, differences are still observable due to the specific amino acid and its immediate neighbors. Moreover, the variation is expected to increase as the bond becomes more extended. To identify the amino acids that produce a shift in the storage of energy compared to others, we show the energies detected in this bond as a function of the amino acid in that position (Fig.~\ref{fig:dataset}c). Among the different degrees of freedom of the backbone, the one storing most of the energy is the C$_\alpha$-C distance, and therefore, we focus on it (see also the discussion about the barriers in SI). However, we observe very similar trends for the other backbone bonds (see Fig. S5 for the analysis of the C$_\alpha$-N and C-N bonds).  Across tripeptides sharing the same central amino acid, we observe a distribution of absorbed energies, with mean values that do not differ significantly among most central amino acid types (Fig. ~\ref{fig:dataset}c). However, we notice that glycine (G) has a tendency to store more energy than the other amino acids, while proline (P) produces the opposite effect: it lowers the amount of energy stored in the C$_\alpha$-C bond. According to the ANOVA test~\cite{heiman_understanding_2001} (Figure~\ref{fig:dataset}d), proline and glycine indeed significantly differ from the other amino acids in the storage of energy of the C$_\alpha$-C bond. We support this result with the F-statistics test (Figure S4), which also indicates that the energies for glycine and proline produce differences that are statistically significant.

The question arises if molecular mechanics force fields yield an energy distribution which is in line with our quantum chemical results from SITH. To answer this question, we built peptides formed by the 21 amino acids organized randomly, and subjected them to reactive molecular mechanics simulations of stretched collagen fibrils using KIMMDY~\cite{rennekamp_hybrid_2020, hartmann_kimmdy_2025}. KIMMDY evaluates the rupture propensities of bonds based on the extent by which a bond is energetically destabilized by the external stretching force. Instead of Pro and Gly, the amino acids Asn, Val, and Meth are more prone to break (Fig. S6), presumably due to the way they are parametrized. When it comes to collagen, the protein for which bond rupture has been most extensively studied up to now, glycine instead has the largest tendency to break, in proportion to its abundance, and proline rupture is completely suppressed (Fig. S7). In this system, importantly, secondary and tertiary structures, including interactions across the polymer chains, influence rupture propensities in addition to the mere local chemical environment of the bond. Our SITH results suggest that the intrinsic differences in energy distribution in glycine and prolines, which is overlooked in current KIMMDY simulations, will likely further boost glycine ruptures and suppress proline ruptures when taken into account in such hybrid kinetic and classical simulations.

\section{Discussion and Conclusion}

SITH models a decomposition of the total change of energy of a molecular system into a distribution of energies in its degrees of freedom using quantum chemical methods. We developed SITH inspired by previous methods, in particular the analysis of forces of Avdoschenko {\it et al.}~\cite{avdoshenko_calculation_2014} and JEDI, developed by Stauch and Dreuw~\cite{stauch_quantitative_2014, stauch_advances_2016, stauch_quantum_2017,wang_jedi_2024}. The novelty of SITH is the numerical integration of the work-energy theorem during a stretching process. In this way, SITH achieves high accuracy, yielding the expected total change of energy computed with DFT (Fig.~\ref{fig:trialanine}c and Fig.~\ref{fig:proline}b).

SITH is limited by definition to sets of a maximum of 3N-6 degrees of freedom. This is a fundamental limitation to guarantee unique variation of each degree of freedom in the partial derivative used to obtain the forces (equation~\ref{eq:force_i}). This implies a current limitation of SITH to be applied to systems highly connected, such as graphene networks. However, SITH is still applicable to a huge diversity of molecules whenever a z-matrix can be defined which contains all relevant degrees of freedom of the force-bearing scaffold. In particular, we show that SITH can shed light on the analysis of the distribution of energies in the bonds of the backbone of peptides, and by extension, to proteins. Even more encouraging, we found robust decomposition of energies by SITH in prolines, and expect SITH to give trustworthy results also when stretching other polymers or mechanophores with individual rings, such as conjugated polymers~\cite{brosz_martini_2022}, or mechanophores with cyclopropane or cyclobutane and their derivatives~\cite{lbrown_molecular_2015}. 

JEDI also has fundamental limitations. In particular, it relies on a harmonic approximation that overestimates the stiffness of the bonds (Fig.~\ref{fig:trialanine}d). These two problems of JEDI can lead to an overestimation of the total change of energy (Fig.~\ref{fig:trialanine}c). While we are aware of the limitations of SITH, we can nevertheless guarantee that our approach is based on a solid framework for prediction and interpretability, even for stretchings far from the energy minimum. We therefore view JEDI and SITH as complementary methods: JEDI can be used to identify the force-bearing scaffold, after which SITH can be applied to study the distribution of energies in stretched molecules up to the point of bond cleavage.

One of the strengths of SITH is that the interpretation of distribution of energies is straightforward: Each value corresponds to an interaction (equation~\ref{eq:interactions}). From this perspective,  SITH can be also used as a benchmark for classical force fields.

SITH shows that peptides, and by transferability also proteins, that are stretched by an external force are more susceptible to break in the C$_\alpha$-C bond of the glycine compared to other bond types and amino acids. On the other side, SITH predicts that prolines stabilizes the C$_\alpha$-C bond, likely by diverting a part of the stretching force into the other bonds of the 5-membered ring. In other words, in a competition scheme, where an external force stretches a peptide, the most likely bond to primarily break is the C$_\alpha$-C bond of the glycine and the least likely would be the C$_\alpha$-C of the proline; with the other C$_\alpha$-C bonds of the other amino acids in between. Naturally, on top of this mere local chemical effect of relative bond stabilities, the local and global structure of the protein, including secondary structure, long-range interactions, and covalent crosslinks, will further modify the energy distribution and rupture propensities. However, the intrinsic rupture propensities determined by the amino acid type can still be influential, but are often overlooked in atomistic and coarse-grained simulations. The same applies to other biological (RNA, DNA)~\cite{hahmann_sequence-specific_2025}  and non-biological heteropolymers ~\cite{willis-fox_polymer_2018, , caruso_mechanically-induced_2009, lee_relative_2015, akbulatov_experimentally_2017}.
We believe that SITH, by revealing internal energy distributions, has the potential to add quantitative insights to complex problems in the field of polymer mechanochemistry.

\section{Data Availability}

All data supporting the findings of this study are publicly available at  \url{https://doi.org/10.17617/3.MW21DN}

\section{Code availability}

SITH sofware ---including documentation and tutorials--- is publicly available at \url{https://graeter-group.github.io/sith/}.

\section{Acknowledgments}

D.S, A. D. and F.G gratefully acknowledge the support of the Klaus Tschira Foundation (SIMPLAIX project 8).
D.S and F.G. gratefully acknowledge the provision of computing resources by the state of Baden–Württemberg through bwHPC and the German Research Foundation (DFG) through Grant Nos. INST 35/1597-1 (Helix cluster). 
This project has received funding from the European Research Council (ERC) under the European Union’s Horizon 2020 research and innovation programme (grant agreement No. 101002812) (to F.G.). We thank Dmitrii E. Makarov and Tim Stauch for fruitful discussions and valuable insights that helped shape this work.
\printbibliography

\newpage
\vspace*{\fill}
{\title{\huge{Supporting Information}}}
\vspace*{\fill}
\newpage

\section{SITH Energy Decomposition for Penta-Alanine.}

\begin{figure}[!h]
    \centering
    \includegraphics[width=\textwidth]{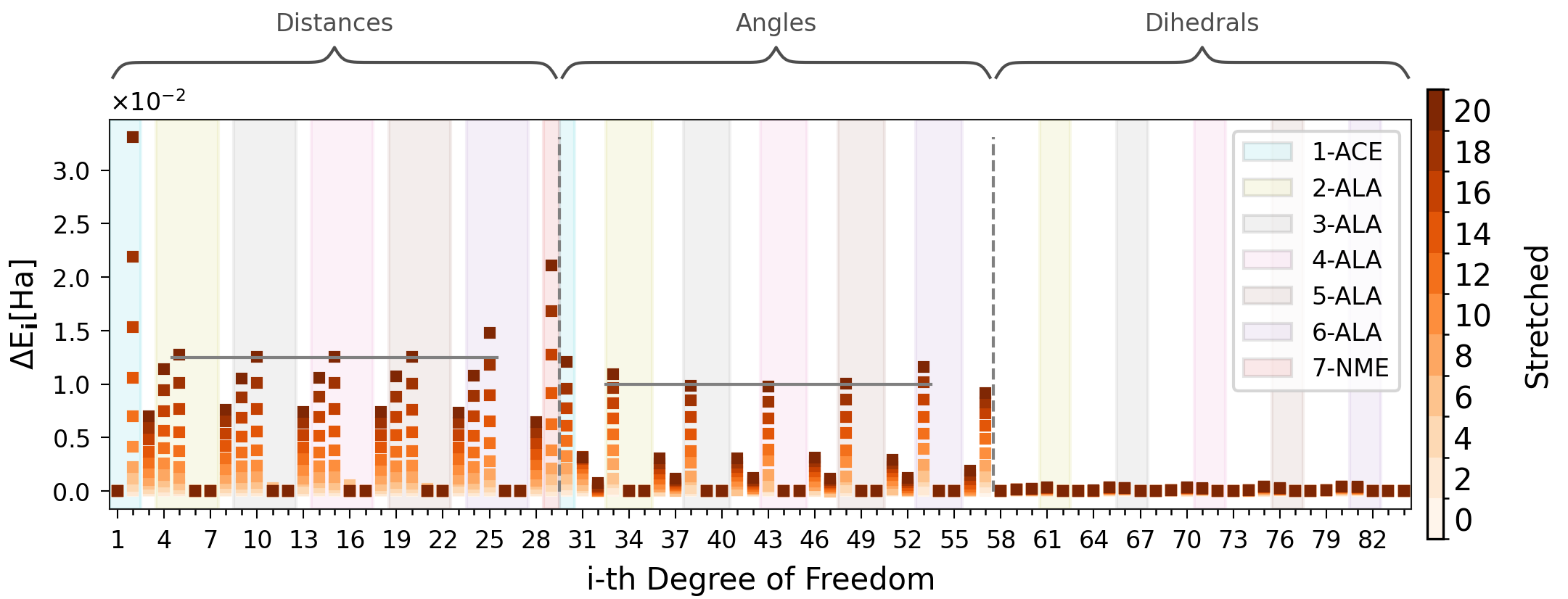}
    \caption{SITH energy decomposition for penta-alanine in each
             degree of freedom as the stretching increases (color bar). The degrees of freedom are
             grouped in distances, angles and diherals, with background colors indicating
             the corresponding residue (white: atoms of the two adjacent residues are involved).The gray horizontal line shows that the energy stored in the three alanines in the middle is the same, while the two alanines at the extremes are biased by the capping groups.}
    \label{fig:abundance}
\end{figure}

\newpage

\section{Difference in Grappa and Amber99 energy Calculation in Respect to SITH Prediction.}

\begin{figure}[!h]
    \centering
    \includegraphics[width=\textwidth]{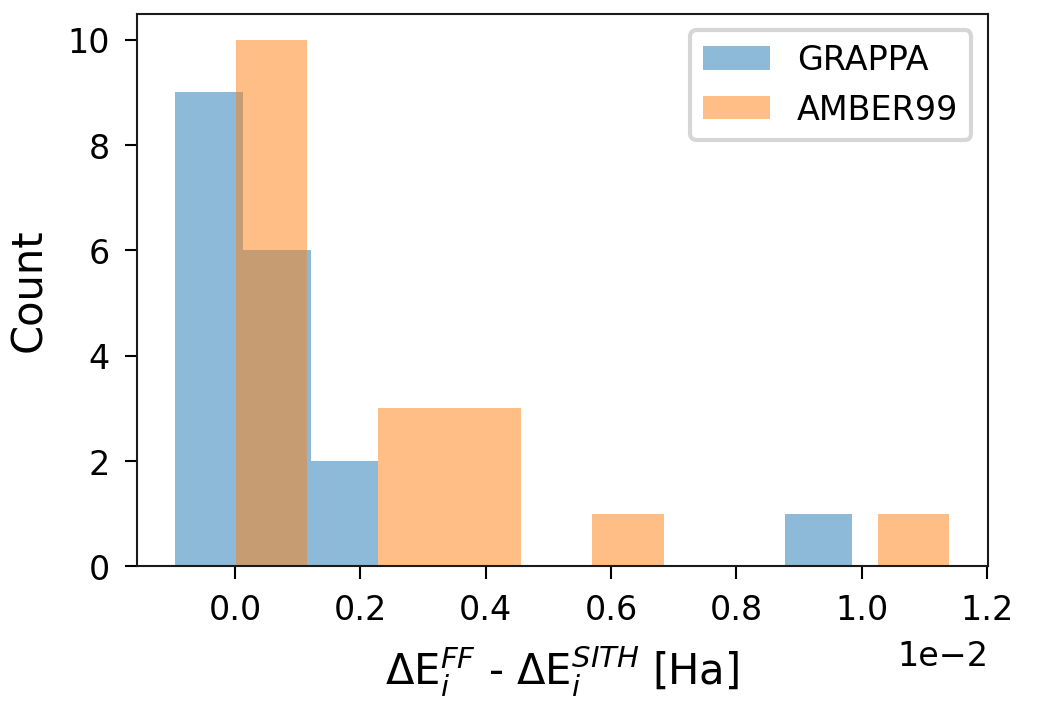}
    \caption{Difference between the energy of the interaction in the degrees of freedom ($\Delta$E$_i$, with i distances, angles and dihedrals) computed by Grappa and Amber99 in comparison with the SITH prediction. Notice that amber differences are all larger than the SITH prediction, and there are more values far from zero. In contrast, some values of Grappa are  smaller than the SITH prediction (which compensates for the positive one in the prediction of the total energy of the system), and the distribution is more centered around zero.}
    \label{fig:abundance}
\end{figure}

\newpage
\section{Frequencies of Amino Acids in the data set.}

\begin{figure}[h!]
    \centering
    \includegraphics[width=\textwidth]{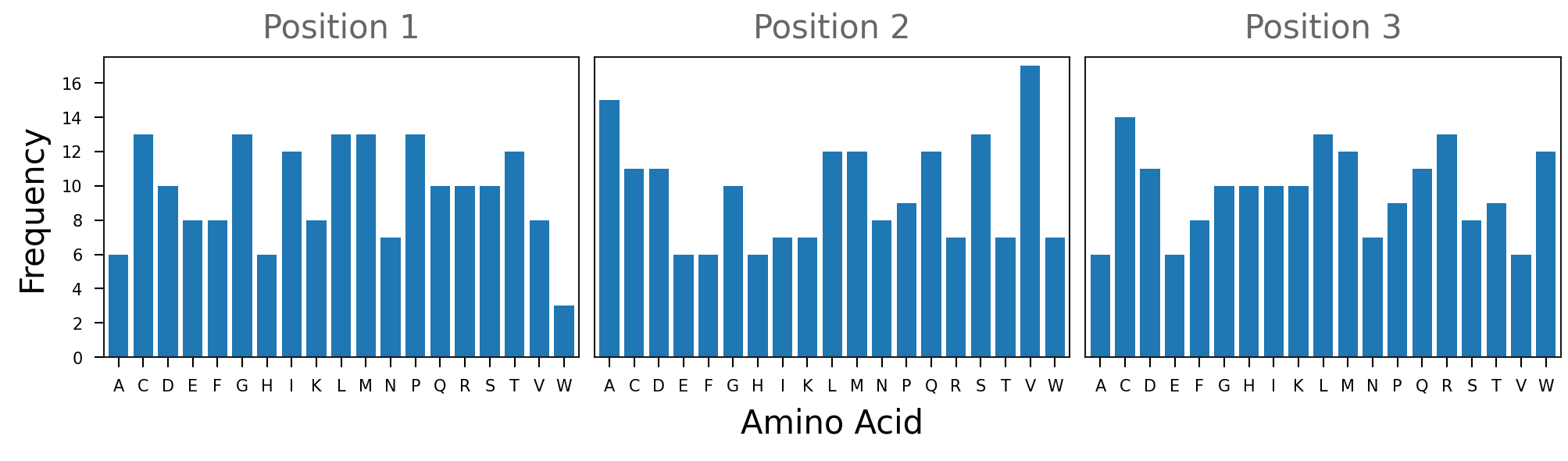}
    \caption{Amount of amino acids in the tripeptides dataset separated by its position in the peptide. Here we use the one-letter-code expanded in Table \ref{table:aminos}.}
    \label{fig:abundance}
\end{figure}

\newpage
\section{One Letter Code for Amino Acids}

\begin{table}[h!]
    \centering
    \caption{One letter code for the 21 amino acids used along this paper.}
    \vspace{0.4cm}
    \begin{tabular}{|c|c|}
        \hline
         One-letter code & Amino Acid  \\
         \hline
         A & Alanine         \\
         C & Cysteine        \\
         D & Aspatric Acid   \\
         E & Glutamic Acid   \\
         F & Phenylalanine   \\
         G & Glycine         \\
         H & Histidine       \\
         I & Isoleucine      \\
         K & Lysine          \\
         L & Leucine         \\
         M & Methiodine      \\
         N & Asparagine      \\
         P & Proline         \\
         O*& Hydroxyproline  \\
         Q & Glutamine       \\
         R & Arginine        \\
         S & Serine          \\
         T & Threonine       \\
         V & Valine          \\
         W & Tryptophan      \\
         Y & Tyrosine        \\
         \hline
    \end{tabular}
    \label{table:aminos}
\end{table}

\newpage
\section{F-test: C$_\alpha$-C Energies are Statistically Different.}

\begin{figure}[!h] \label{fig:f-test}
    \centering
    \includegraphics[width=\textwidth]{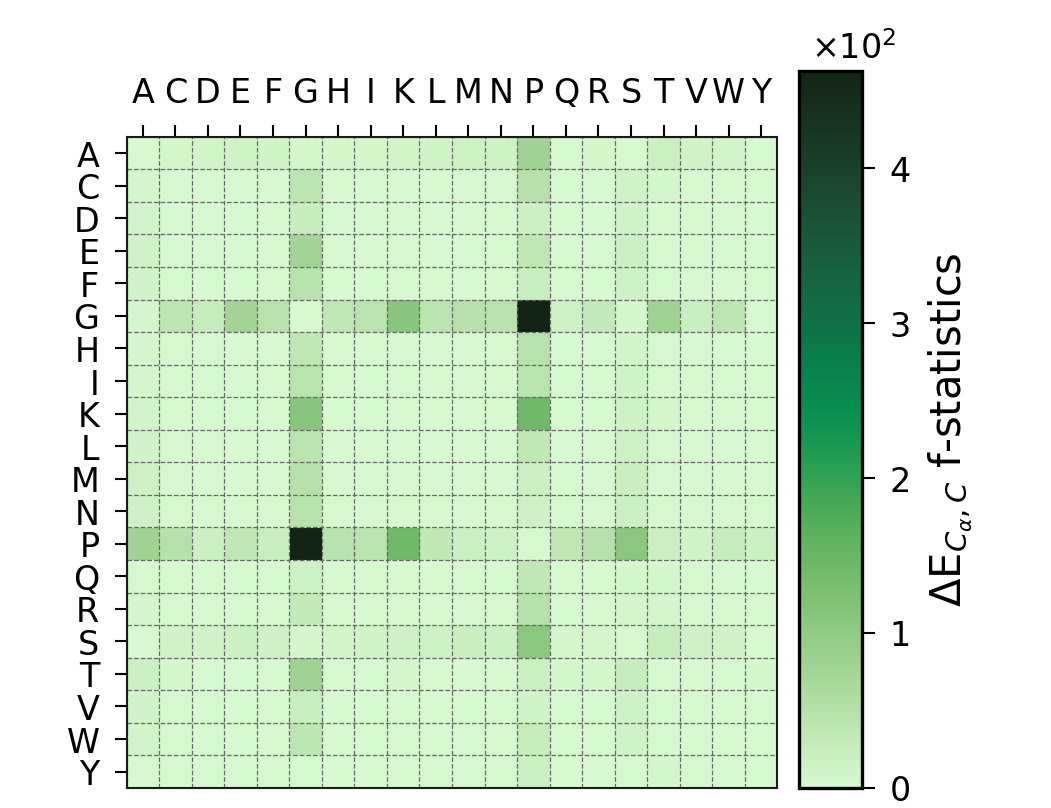}
    \caption{F-statistic test between the SITH energy predictions of the C$_{\alpha}$-C bonds of the tri-peptides of the dataset. The higher the f-statistics value, the larger the confidence that the variances differ more than random noise~\cite{heiman_understanding_2001}. The C$_{\alpha}$-C bonds considered here are the ones of the amino acids in the middle of the peptides, shown in the axis as one-letter-code expanded in Table ~\ref{table:aminos}. Notice that the rows and columns of the proline and glycine are darker than the rest of the amino acids. Especially the f-value between the glycine and proline distributions.}
\end{figure}

\newpage

\section{Energy Stored in C$_\alpha$-C and C$_\alpha$-N bonds}

\begin{figure}[!h]
    \centering
    \includegraphics[width=\textwidth]{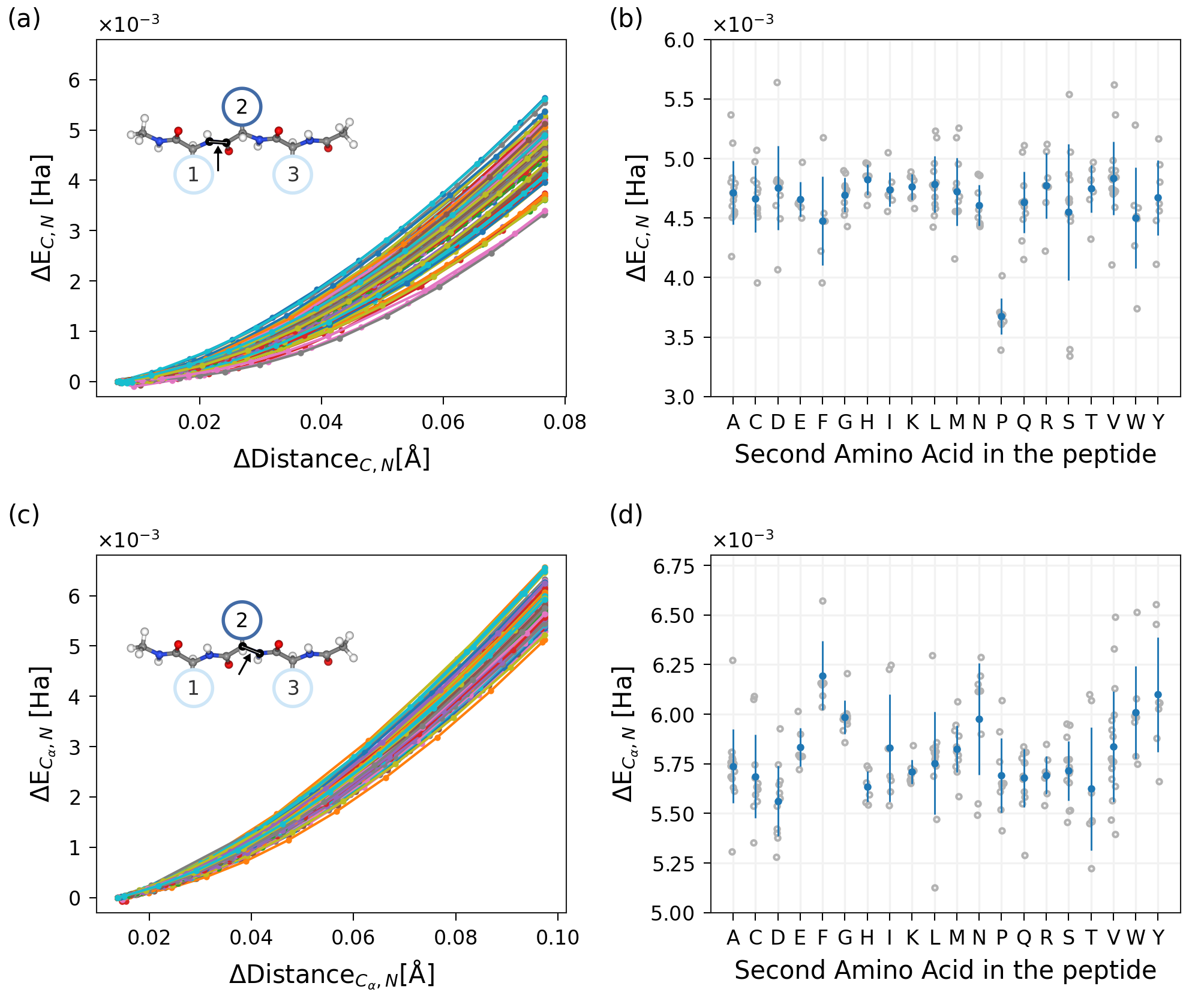}
    \caption{a), c) Energy stored in the C–N and C$_\alpha$–N bonds as a function of bond extension; each curve corresponds to a different tripeptide. The stretched bond is illustrated on the tripeptide, where labels 1, 2, and 3 denote the first, second, and third amino acids. b) Distribution of energies stored in the C–N bond across the peptide dataset after stretching the bond by 0.078 Å from its equilibrium distance, shown as a function of the identity of the middle amino acid. d) Distribution of energies stored in the C$_\alpha$–N bond across the peptide dataset after stretching the bond by 0.97 Å from its equilibrium distance, shown as a function of the identity of the middle amino acid. In b) and d), gray points represent individual peptides with different combinations of amino acids at positions 1 and 3; blue dots indicate the mean values, and error bars denote the standard deviations.}
    \label{fig:CaN_CN}
\end{figure}

\newpage

\section{KIMMDY Rupture of Polypeptides}

We obtain ruptures in polypeptides using KIMMDY. We run KIMMDY in four peptides formed by 21 amino acids organized randomly and duplicated. The four peptides are the following in the one-letter-code presented in Table~\ref{table:aminos}:

\begin{center}
    ARNDCEQGHILKMFO*PSTWYVARNDCEQGHILKMFO*PSTWYV
    HLMPO*TYANCQIKFSWVRDEGHLMPO*TYANCQIKFSWVRDEG
    YKMPFATQVDIRSEHCWNO*LGYKMPFATQVDIRSEHCWNO*LG
    YTHCDKEO*QPMLRFSGVIAWNYTHCDKEO*QPMLRFSGVIAWN
\end{center}

\begin{figure}
    \centering
    \includegraphics[width=\textwidth]{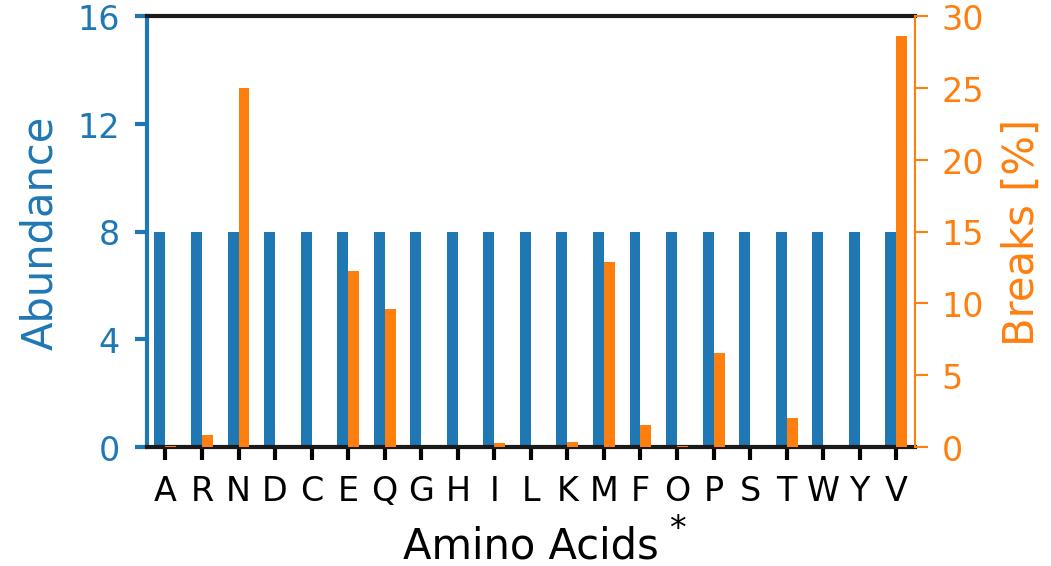}
    \caption{Abundance and percentage of breaks of amino acids in polypeptides pulled by an external force of 1~nN and 1.5~nN. The amino acids are presented in Table \ref{table:aminos}.}
    \label{fig:polypep_rupture}
\end{figure}

We pull each one of the polypeptides with 1~nN and 1.5~nN. Then, we generate 500 bond ruptures per simulation according to the KIMMDY workflow. In Figure~\ref{fig:polypep_rupture}, we show that even when the population of each amino acid in all the peptides is the same, there are some amino acids that are way more likely to present a rupture than others. This bias does not correspond to our results with SITH.

\newpage

\section{KIMMDY Rupture in Collagen}

\begin{figure}[!h]
    \centering
    \includegraphics[width=\textwidth]{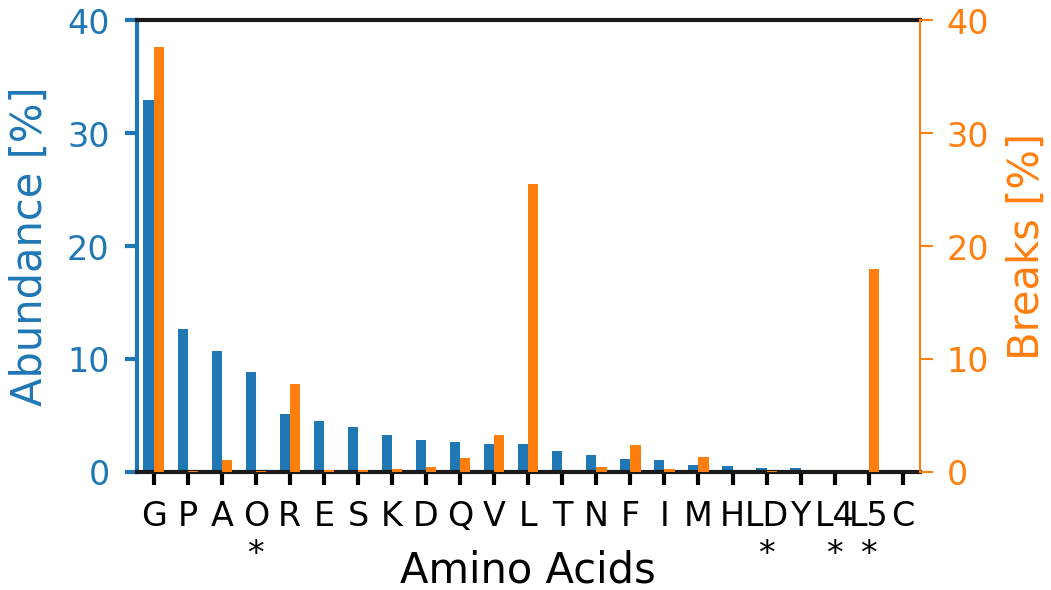}
    \caption{Percentage of abundance and breaks of amino acids in collagen fibrils pulled by an external force. The amino acids are presented in Table \ref{table:aminos}. The amino acids with asterisks are non-natural amino acids: O is hydroxyproline, LD is DOPA, L4 and L5 are crosslinks.}
    \label{fig:abundance}
\end{figure}

\newpage
\section{Barrier Change as Stretching Increases}

\begin{figure}
    \centering
    \includegraphics[width=\textwidth]{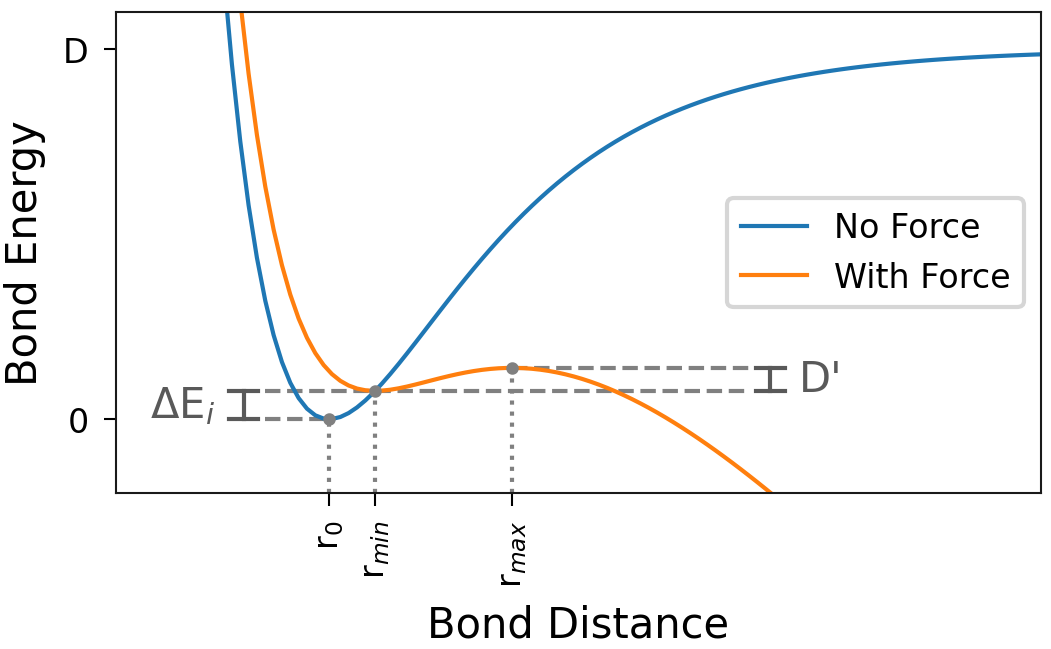}
    \caption{Bond energy vs bond distance according to Bell-Evans model. The blue curve is the bond interaction with no external force, described as a Morse potential with a minimum at $r_0$ and a dissociation energy $D$. The orange curve is the bond interaction after an external force is applied. The external force increases the equilibrium distance to $r_{min}$, changes the location of the barrier to $r_{max}$ and decreases the dissociation energy from $D$ to $D'$. We show the energy in the bond, $\Delta E_i$, computed using SITH.}
    \label{fig:bondinteraction}
\end{figure}

Using the Bell-Evans model~\cite{bell_models_1978}, we describe the bond interaction after the influence of an external force, $F_i$, as

\begin{equation}\label{eq:interaction}
    e_i(r) = D (1 - e^{-\beta( r-r_0)})^2 - F_ix~,
\end{equation}

\noindent
where $i$ is the label of the bond, $D$ is the well depth (dissociation energy), $\beta$ is a constant, and $r_0$ is the equilibrium distance of the bond when no force is applied (Fig. \ref{fig:bondinteraction}). In mechanical equilibrium, the force $F_i$ can be written in terms of the energy stored in the degree of freedom computed by SITH, $\Delta E_i$, as

\begin{equation}
    F_i = 2\beta({\sqrt{D\Delta E_i} - \Delta E_i})~.
\end{equation}
\noindent
Notice that here we assume that the energy in the bond obtained with DFT level of accuracy using SITH is well described by a morse potential. This is one of the hypotheses of the Bell-Evans model.

The external force in equation \ref{eq:interaction} increases the distance of minimum energy, changes the position of the transition between the bonded and unbonded states, and decreases the dissociation energy (Fig. \ref{fig:bondinteraction}). The value of the new barrier is given by

\begin{equation}\label{eq:delta_bar}
    D' = e_i(r_{max}) - e_i(r_{min})~,
\end{equation}

\noindent
where $r_{min}$ and $r_{max}$ are the lengths of the new minimum and maximum, respectively, which are given by

\begin{equation}\label{eq:minmax}
    r_{min/max} =  r_0 - \frac{1}{\beta}\textbf{ln}\left( \frac{1}{2} \pm {\frac{1}{2}} \sqrt{\frac{\beta D - 4\beta (\sqrt{D \Delta E_i} - \Delta E_i)}{\beta D}}\right)~.
\end{equation}

Replacing (\ref{eq:interaction}) and (\ref{eq:minmax}) into (\ref{eq:delta_bar}), the new barrier in terms of the energy stored in the bond as predicted by SITH is

\begin{equation}\label{eq:newbarrier}
\begin{split}
    D' = & \hspace{2mm} D\sqrt{{\frac{-4\sqrt{ D\Delta E_i} +  4\Delta E_i + D}{D}}} \hspace{3mm}+ \\
         & \hspace{2mm} 4 (\Delta E_i - \sqrt{D \Delta E_i}) \hspace{2mm}\textbf{tanh}^{-1}\left(\sqrt{\frac{-4\sqrt{D\Delta E_i} + 4\Delta E_i +D}{D}}\right)~.
\end{split}
\end{equation}

\begin{figure}
    \centering
    \includegraphics[width=\textwidth]{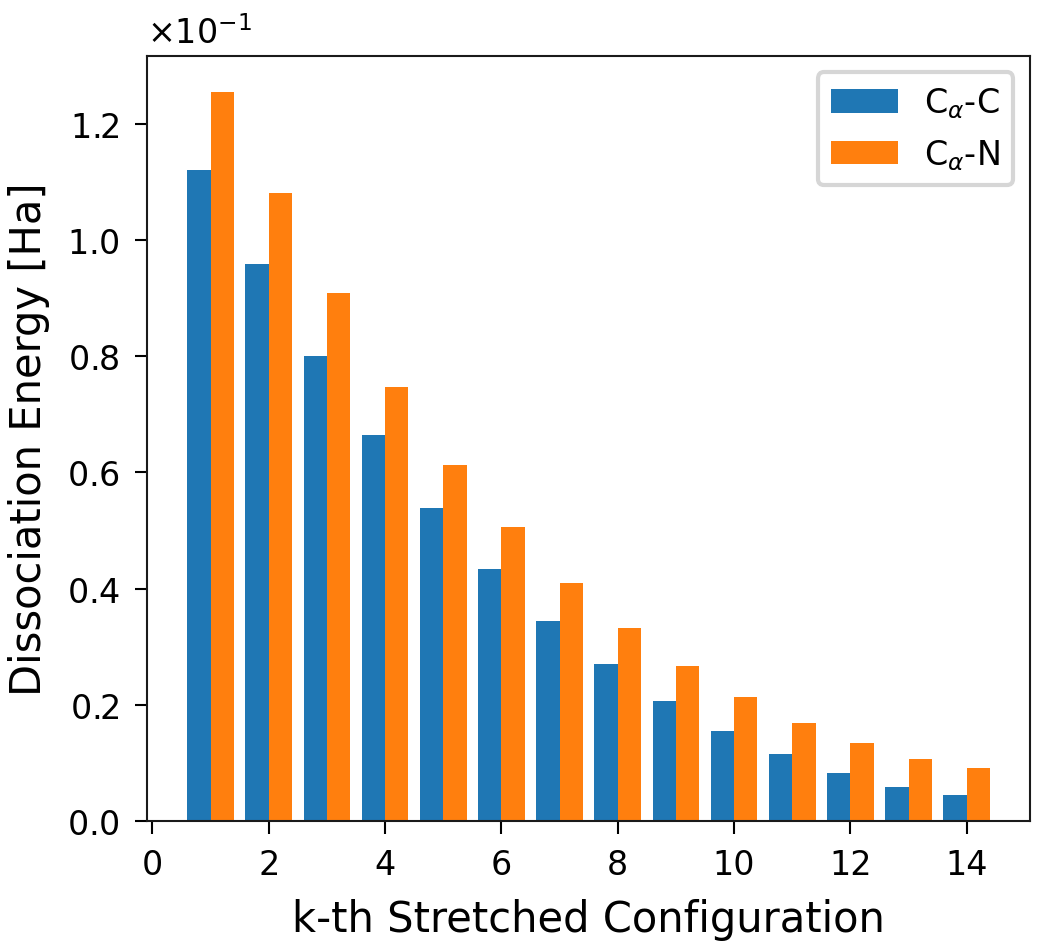}
    \caption{Dissociation energy of the C$_\alpha$-C and C$_\alpha$-N bonds of the middle alanine in a tri-alanine peptide as the end-to-end stretching increases.}
    \label{fig:barrierChange}
\end{figure}

Using this result, we prove that in stretching conditions, the C$_\alpha$-C bond is weaker than the C$_\alpha$-N bond. We compute the new dissociation barrier for the C$_\alpha$-C and C$_\alpha$-N for the alanine in the middle of a trialanine peptide. We take the values of the bond dissociation energies, $D$, presented by Rennekamp {\it et al.} \cite{rennekamp_collagen_2023}, which are 0.130 Ha and 0.144 Ha for the C$_\alpha$-C and C$_\alpha$-N respectively. Then, we compute the final barrier from equation \ref{eq:newbarrier} (Fig. \ref{fig:barrierChange}). We notice that the barrier of the C$_\alpha$-C bond is always lower than the C$_\alpha$-N bond. At the maximum stretched configuration, the relative probability of rupture of the C$_\alpha$-C bond is 1.6 times higher than the probability of rupture of the C$_\alpha$-N bond. For this reason, we focus on the analysis of the C$_\alpha$-C bond in the main text.

\end{document}